\documentclass[12pt]{article}

\usepackage{amssymb}

\usepackage{verbatim}

\def\bR {{\mathbb{R}}}
\def\bN {{\mathbb{N}}}

\def\bH {{\mathbb{H}}}

\def\bZ {{\mathbb{Z}}}

\def\cK {{\mathcal K}}

\def\cN {{\mathcal N}} 
\def\cO {{\mathcal O}}

\def\Di {\displaystyle}

\def\MR {\mathrm }

\def\pR {{\mathbb R}\times {\mathbb R}^*_+}
\def\sp {\mathrm {sp}}
\def\mb {{\bf b}} 
\def\mbt {{\widetilde {\bf b}}}

\makeatletter
\@addtoreset{equation}{section}

\makeatother

\newtheorem{theorem}{Theorem}[section]
\newtheorem{lemma}[theorem]{Lemma}
\newtheorem{proposition}[theorem]{Proposition}

\newtheorem{remark}[theorem]{Remark}
\newtheorem{corollary}[theorem]{Corollary}

\begin{document}

\bibliographystyle{plain}

\begin{center}
\Large \bf {
 Magnetic bottles on the Poincar\'e half-plane: spectral 
 asymptotics}
\end{center}

\vskip 0.5cm

 \centerline {\bf { Abderemane MORAME$^{1}$
 and  Fran{\c c}oise TRUC$^{2}$}}

{\it {$^{1}$ Universit\'e de Nantes,
Facult\'e des Sciences,  Dpt. Math\'ematiques, \\
UMR 6629 du CNRS, B.P. 99208, 44322 Nantes Cedex 3, (FRANCE), \\
E.Mail: morame@math.univ-nantes.fr}}

{\it {$^{2}$ Universit\'e de Grenoble I, Institut Fourier,\\
            UMR 5582 CNRS-UJF,
            B.P. 74,\\
 38402 St Martin d'H\`eres Cedex, (France), \\
E.Mail: Francoise.Truc@ujf-grenoble.fr }}

 \footnote{ {\sl Keywords}~: spectral asymptotics, 
  magnetic bottles, hyperbolic plane
 , minimax principle.}

\begin{abstract}
We consider a  magnetic Laplacian  
 $-\Delta_A=(id+A)^\star (id+A)$\\  
 on the Poincar\'e upper-half plane $\bH \; ,$ 
 when the magnetic field $dA$ is infinite at the infinity such that 
  $-\Delta_A$ has pure discret spectrum. 
  We give the asymptotic behavior of the counting function 
  of the eigenvalues. 
\end{abstract}

\section{Introduction}

In this paper we study the asymptotic distribution of large eigenvalues
of  magnetic bottles  on the hyperbolic plane $\bH$.
Magnetic bottles on $\bH$ are Schr\"odinger operators of the form
\begin{equation}\label{def1H}
-\Delta_A\; =\; y^2(D_x-A_1)^2\; +\; y^2(D_y-A_2)^2\; ,
\end{equation} 
where the magnetic field  $dA$ is infinite at the infinity . This 
property ensures that  $-\Delta_A$ has a compact resolvent.
 The precise formulation is given below.

  In the Euclidean case the asymptotic distribution of large eigenvalues
of  magnetic bottles in $\bR^d$ 
has been given by Yves Colin de Verdi\`ere \cite{Col},
using partition in cubes and estimations for constant magnetic fields in 
the cubes. This method can still be used here, but
cubes are replaced by rectangles adapted to the hyperbolic geometry 
and the formula we get is of the same type, taking into account the
hyperbolic volume and the hyperbolic definition for magnetic fields.

 The hyperbolic framework we recall below has been used mainly for
studying the Maass Laplacian , which corresponds to the constant magnetic
field case. This case has been studied by  many authors (see 
\cite{Gro}, \cite{Els}, \cite{Com}   \cite{D-I-M}). In 
\cite {In-Sh1}   the authors consider asymptotically constant
 magnetic fields  and in \cite {In-Sh3} they deal with
 Pauli operators. See also \cite {Ike} for relationship between Maass Laplacian   
and Schr\"odinger operators with Morse potentials.

From an other point of view , the asymptotic distribution of large eigenvalues
in the  hyperbolic context has already been studied for Schr\"odinger
operators (without magnetic field) (see \cite {In-Sh2}) . The method is based on
 Feynman-Kac representation of the heat kernel and the Tauberian theorem. As already mentioned our own
method involves only min-max techniques so it does not require to study  
properties of the evolution semigroup. It is also local, so our result is valid
for many surfaces of infinite area with fundamental domain $\bH$ .

  Let us now set up the hyperbolic framework of our problem.

In a connected and oriented Riemannian manifold 
$(M, g)$ of dimension $n\; ,$ for any real 
 one-form $A$ on $M\; ,$ one can define the magnetic Laplacian 
\begin{equation}\label{DeA} 
\begin{array}{c} 
- \Delta_A\; =\; (i\ d + A)^\star (i\ d +A)\; ,\\ 
 \left ( \; 
 (i\ d+A)u=i\ du+uA\; , \ \forall \; u\; \in \; C^\infty_0(M)\; \right )\; . 
 \end{array} 
 \end{equation} 
The magnetic field is the exact two-form 
$\ \rho_B\; =\; dA\; .$\\ 
The two-form $\rho_B$ is associated with a linear operator $B$  
on the tangent space  defined by 
\begin{equation}\label{defB} 
\rho_B(X,Y)\; =\; g(B.X,Y)\; ;\quad \forall \; X\; ,\ Y\; \in TM \times TM\; .
\end{equation}
The magnetic intensity $\mb$ is given by
\begin{equation}\label{defb} 
\mb\; =\; \frac{1}{2} tr\left ( (B^\star B)^{1/2} \right )\; . 
\end{equation}

Let us assume that $dim(M)=2 ,$ and denote by  $dv\; $  the Riemannian measure on $M$; then 
$\rho_B\; =\; \mbt \; dv\; ,$ 
 with $\; |\mbt |\; =\; \mb \; .$ \\ 
In this case, we can say that the magnetic field is constant iff 
$\mbt $ is constant.

Now, we consider the case where $M\; =\; \bH \; $ is the hyperbolic plane~: 
$$\bH\; =\; \bR \times ]0,+\infty [\; ,\quad g\; =\; \frac{dx^2+dy^2}{y^2}\; ,
\quad A = A_1(x,y)\ dx\; +\; A_2(x,y)\ dy\; .$$ 
We will assume that 
\begin{equation}\label{hypA}
 A_j(x,y)\; \in \; C^2(\bH ;\bR )\; ,\quad \forall \; j\; . 
\end{equation} 
Let us define $D_x = \frac{1}{i} \partial_x $ and $D_y = \frac{1}{i} \partial_y $ .
Then we have 
\begin{equation}\label{def1H}
-\Delta_A\; =\; y^2(D_x-A_1)^2\; +\; y^2(D_y-A_2)^2\; ,
\end{equation} 
$$ 
\mbt \; =\; y^2\left ( \partial_xA_2 -\partial_yA_1\right ) \quad \mb 
\; = \; |\mbt |\; ,\quad 
\MR{and}\quad dv\; =\; y^{-2}dxdy\; .
$$  

It is well known that  
$\; -\Delta_A\; $  
is essentially self-adjoint on $L^2(\bH)\; ,$ 
see for example \cite{Shu}.

 As we are only interested on the spectrum of $\;  \sp (-\Delta_A)$,  
we will  use that it is gauge invariant: 
\begin{equation}\label{gauge} 
\sp (-\Delta_A)\; =\; \sp (-\Delta_{A+d\varphi })\; ;
\quad \forall \; \varphi \; \in \; C^2 (\bH ; \bR )\; .
\end{equation} 

For an operator $H\; ,\ \sp (H)\; ,\ \sp_{es}(H)\; ,\ \sp_a(H)\; ,
\ \sp_d(H)\; $ and $ \sp_p(H)\; $ denote its spectrum, 
its essential part, its absolutely continuous part, its discret part 
and its ponctual part.

We will denote $\ -\Delta_A\; $ by $\ P(A)\; .$

\section{The  constant magnetic Laplacian on the hyperbolic plane} 
      
In this section, we explain how to get the well-known properties 
of the spectrum of a constant magnetic Laplacian on the hyperbolic 
plane.  The original study was done by J. Elstrodt in \cite{Els}.   
      
We consider the case where 
$\ y^2 (\partial_x A_2(x,y) - \partial _yA_1(x,y))$ is constant.  
We choose a gauge such that $A_2=0\; ,$ so $\ A_1(x,y)=\pm \mb y^{-1}\; .$ 
We can assume that $\ A_1(x,y)= \mb y^{-1}\; ,$ even if we change $x$ 
into $-x\; ,$ which is a unitary operator on $L^2(\bH )\; .$ 
The operator we are interested in is 
\begin{equation}\label{DeltaBC} 
-\Delta_{A^\mb } \; =\; y^2(D_x-\mb y^{-1})^2\; +\; y^2D_y^2\; ,\quad 
\MR{with}\quad \mb\geq 0\quad \MR{constant.} 
\end{equation} 
Let $U$ be the unitary operator 
\begin{equation}\label{defU} 
U\; :\; L^2(\bH )\; \to \; L^2(\pR )\; ,\quad 
Uf\; =\; y^{-1}f\; ; 
\end{equation} 
$\pR $ is endowed with the standard Lebesgue measure $dxdy\; .$ 
Then 
\begin{equation}\label{defPb} 
P_\mb\; =\; U(-\Delta_{A^\mb })U^\star \; =\; (D_x-\mb y^{-1})y^2(D_x-\mb y^{-1})\; 
+\; D_y y^2 D_y\; .
\end{equation} 

Using partial Fourier transform we get that 
$\Di \sp (P_\mb )\; =\; \bigcup_{\xi \in \bR } \sp (P_\mb (\xi ))\; ,$ 
where $P_\mb (\xi )$ is the self-adjoint operator 
on $L^2(\bR_+)$ defined by 
\begin{equation}\label{defPxi}
 P_\mb (\xi )f\; =\; (y\xi -\mb )^2f(y) +D_y (y^2 D_yf )(y)  \; ; 
\quad \forall \; f\; \in \; C^{\infty}_{0}(\bR_+)\; . 
\end{equation} 
When $\xi >0\; ,$ by scaling, $\ y\; \to \xi^{-1}y\; ,$ we get that 
$$\sp (P_\mb (\xi ))\; =\; \sp (P_\mb (1))\; ,\quad (\MR{if}\quad \xi >0)\; .$$


In the same way, we get that 
$$\sp (P_\mb (\xi ))\; =\; \sp (P_\mb (-1))\; ,\quad \MR{if} \quad 
\xi \; <\; 0\; .$$ 
It is easy to see that $\ \sp _{es} (P_\mb (\pm 1))\; 
=\; \mb^2\; +\; \sp _{ac} (P_0(1))\; =\; \sp_{ac}(P_\mb (\pm 1))\; ,$ 
\\ 
and,  (see for example the exercise  I6 p. 1573 in \cite{Du-Sc}),   
\begin{equation}\label{Dunford} 
\sp (P_\mb (-1))\; =\; \sp_{ac}(P_\mb (-1))\; =\; 
[\mb^2 + \frac{1}{4} , +\infty [\; =\; 
\sp_{ac} (P_\mb (1))\; . 
\end{equation} 
$P_\mb (1)\ $ may have some eigenvalues in $\Di \ [\mb , \mb^2+\frac{1}{4}[\; .$\\ 
For the proof, we use the method of \cite{In-Sh1}. We define 
\begin{equation}\label{defKb} 
K_\mb\; =\; y-\mb -1 -iyD_y\; ;\quad \MR{so}\quad K_\mb^\star\; =\; y-\mb +iyD_y\; . 
\end{equation} 
Then 
\begin{equation}\label{KbEquat} 
K_\mb^\star K_\mb \; =\; P_\mb (1)\; +\; \mb \quad \MR{and}\quad 
K_\mb K_\mb^\star \; =\; P_{\mb +1}(1)\; -\; \mb -1\; .
\end{equation} 
When $\mb >1/2\; ,$ we  define 
\begin{equation}\label{defPhib} 
\varphi_\mb(y)\; =\; \frac{2^{\mb-1/2}}{\sqrt{\Gamma(2\mb -1)}}y^{\mb -1}e^{-y}\; ,
\quad 
\left( \varphi_\mb \; \in \; \MR{Ker}(K_{\mb -1}^{\star}) \right)\; , 
\end{equation} 
$\varphi_\mb$ is the ground state of $P_\mb (1)\; :\quad 
P_\mb (1)\varphi_\mb \; =\; \mb \varphi_\mb \; .$  
\\ 
As $\Di \ K_\mb \left ( P_\mb (1)+2\mb +1 \right )\; =\; P_{\mb +1}(1) K_\mb $\\ 
and $\Di K_{\mb}^{-1}f(y)\; =\; y^{-\mb -1}e^y\int_{y}^{+\infty} s^\mb e^{-s}f(s)ds\; ;\quad 
\forall \; f\; \in \; [\varphi_{\mb +1}]^{\perp}\; ;$\\ 
we get that,  if $\Di \mu+2\mb +1<\mb^2+\frac{1}{4}\; ,$ then 
$$ \mu \; \in\; \sp_d(P_\mb (1))\; \Rightarrow \; \mu+2\mb +1\;
 \in \; \sp_d(P_{\mb +1}(1))\; ,$$   
and if $\ \lambda -2\mb -1\geq \mb \; ,$  
$$ \ \lambda \; \in \; \sp_d(P_{\mb +1}(1))\setminus \{ \mb +1\} 
\; \Rightarrow \; \lambda -2\mb -1\; \in \; \sp_d(P_\mb )\; .$$   
One gets the well-known following theorem: 
\begin{theorem}\label{thPb1} 
The spectrum of $P_\mb (\pm 1)$ is formed by its absolutely continuous part 
and its discret part, and  
\begin{eqnarray*}
\sp (P_\mb (-1))\;& =&\; \sp_{ac}(P_\mb (-1))\; =\; \sp_{ac}(P_\mb (1))\; =\; 
[\mb^2+\frac{1}{4}, +\infty [ \nonumber \\ 
 \sp (P_\mb (1))\;& =&\; \sp_{ac}(P_\mb (1))\; \ \  ,\ \ if \Di \ \mb\; \leq \; \frac{1}{2} \nonumber \\
 \sp_d(P_\mb (1))\;& =&\; \{ (2j+1)\mb -j(j+1)\; ;\ j\; \in \; \bN \; ,\ j<\mb -\frac{1}{2}\} \;\ \   if \Di \ \mb\; >\; \frac{1}{2}\;  .  
\end{eqnarray*}
\end{theorem}

\begin{corollary}\label{spDeltaB} 
The spectrum of $-\Delta_{A^\mb }\; $ is essential: 
$\sp (-\Delta_{A^\mb}) = \sp_{es}(-\Delta_{A^\mb})\; .$\\ 
Its   absolutely continuous part is given by  
$\ \sp_{ac}(-\Delta_{A^\mb} )\; =\; 
[\mb^2+\frac{1}{4}, +\infty [\; .$\\ 
The remaining part of its spectrum is empty if $0\leq  \mb  \leq 1/2\; ,$ 
otherwise it is formed by a finite number of eigenvalues of infinite multiplicity 
given by  
$$   \sp_p(-\Delta_{A^\mb} )\; =\; 
\{ (2j+1)\mb -j(j+1)\; ;\ j\; \in \; \bN \; ,\ j<\mb -\frac{1}{2}\} 
\; ,\quad (\MR{if}\quad \frac{1}{2}<\mb \; .)$$   
\end{corollary}

\section{The case of a magnetic bottle (with compact resolvent)} 
The following theorem deals with the case of a magnetic field
which fulfills magnetic bottles type assumptions. 

\begin{theorem}\label{compactR} 
Under the assumptions (\ref{hypA}) and (\ref{def1H}),  if 
\begin{equation}\label{BVifini} 
\mb (x,y)  \; \to \; +\infty 
\quad \MR{as}\quad d(x,y)\; \to \; +\infty \; , 
\end{equation} 
and if $\exists \; C_0 >0\; $ 
such that, for any vector field $X$ on $\bH \; ,$ 
\begin{equation}\label{Bcontrol} 
|X\mbt | \leq C_0 (|\mbt |+1){\sqrt {g(X,X)}}\; ;  
\end{equation} 
then $P(A)=-\Delta_A \; $ has a compact resolvent. \\ 
$(d(x,y) $ denotes the hyperbolic distance of $(x,y)$ to $(0,1)\; )\ . $ 
\end{theorem} 
{\bf Proof}~: The standard proof for elliptic operators  
on the flat $\bR^n$ can be applied using the estimate given by 
the following Lemma. 

\begin{lemma}\label{MajorB} For any $\epsilon \in ] 0, 1[\; ,$ 
there exists $C_\epsilon >0$ s.t.
 $$\forall \;  f\; \in \; C_{0}^{\infty}(\bH )\; ,
 \quad  \int_{\bH} \mb |f|^2 dv \; \leq \; (1+\frac{\epsilon}{2})
\langle -\Delta_A f | f \rangle_{L^2(\bH )}  + 
C_\epsilon\| f\| _{L^2(\bH )}\; .$$
\end{lemma}
 For the proof,  one can use 
  the unitary operator defined in \ref{defU}

$\Di U\; :\; L^2(\bH )\; \to \; L^2(\pR )\; ,\quad Uf(x,y)=y^{-1}f(x,y)\; .$
 \\ 
We get that\\ 
$\Di UP(A)U^\star  = y^2(D_x-A_1)^2 + y(D_y -A_2)^2y 
\; .$\\ 
In this form, we can write
$\Di UP(A)U^\star \; =\; K^\star K +\mbt =\widetilde{K}^\star \widetilde{K}-\mbt$\\ 
with $\ K\; =\; y(D_x-A_1)\; -\; i(D_y-A_2)y\; $ 
and $\; \widetilde{K}\; =\; y(D_x-A_1)\; +\; i(D_y-A_2)y\; .$ So 
$$\pm \mbt \; \leq \;  UP(A)U^\star \; .$$   
We cover $\pR $ by two open sets $\cO_0,\ \cO_1\; ,$ such 
that $\cO_0$ is bounded and $y$ and $1/y$ are bounded on $\cO_0\; , $ 
and   $1 \leq  \mb $ on $\cO_1\; .$

Taking an associated partition of unity $\chi_j\; ,\ ( j=0,1)\; ,$ 
and using that $\pm \mbt \leq UP(A)U^\star \; ,$ we get  
$$\int_{\pR} \mb |\chi_1f|^2dx dy \; \leq \; \int_{\pR} UP(A)U^\star (\chi_1f)
\overline{\chi_1f} dx dy\; .$$   
The Lemma comes easily from this estimate.

\section{Spectral asymptotics for magnetic bottles}
 
\subsection{The main theorem} 

For a self-adjoint  operator $P\; ,$ and for any 
real $\lambda \leq \inf \sp_{es}(P)\; ,$ 
we denote by $N(\lambda ; P)\; $ the number of eigenvalues of 
$P\; ,$ (counted with their multiplicity), which are in $]-\infty , \lambda [\; .$ 

\begin{theorem}\label{magn} 
Under the assumptions of Theorem \ref{compactR}, 
  $\ P(A)=-\Delta_A \; $ has a compact resolvent and for any 
  
$\delta \; \in \; ]\frac{1}{3}, \frac{2}{5}[\; ,$ in (\ref{HypAD}), 
 there exists a constant $\; C\; > \; 0\; $ such that 
 $$ 
 \frac{1}{2\pi} \int_{\bH} (1-\frac{C}{(\mb (m)+1)^{(2-5\delta)/2}}) 
 \mb (m) \sum_{k=0}^{+\infty} [\lambda (1-C\lambda^{-3\delta +1})- \frac{1}{4}
 - (2k+1) \mb (m)]_{+}^{0} \; dv \; 
 $$  
 \begin{equation}\label{magnE} 
 \leq \; N(\lambda , -\Delta_A)\; \leq   
 \end{equation} 
 $$  
  \frac{1}{2\pi} \int_{\bH} (1+\frac{C}{(\mb (m)+1)^{(2-5\delta)/2}}) 
 \mb (m) \sum_{k=0}^{+\infty} [\lambda (1+C\lambda^{-3\delta +1}) -\frac{1}{4}
 - (2k+1) \mb (m)]_{+}^{0} \; dv \;  
$$
$[ \rho ]^{0}_{+}$ is the Heaviside function: 
$$[ \rho ]^{0}_{+}\; =\; \left\{ \begin{array}{ccc} 1\; ,& \MR{if} & \rho > 0\\ 
0\; , & \MR{if} & \rho \leq 0\; .
\end{array} 
\right . $$ 
\end{theorem} 

This result can be compared to the one obtained in \cite{Col} . The difference between the two results is the additional term $- \frac{1}{4}$ , which comes from the geometry of the problem . It becomes really significant in the
following

\begin{corollary}\label{magnC} 
Under the assumptions of Theorem \ref{compactR} 
and if the function 
$$\omega (\mu )\; =\; \int_\bH  [\mu - \mb (m)]_+^0 dv
$$ 
satisfies 
\begin{equation}\label{hypW} 
\exists \; C_1 > 0\ \MR{s.t.}\ \forall \; \mu > C_1\; ,\ 
\forall \; \tau \; \in \; ]0,1[\; ,\quad 
\omega\ ( (1+\tau )\ \mu) - \omega (\mu ) \leq   C_1 \ \tau\ \omega (\mu ) \; , 
\end{equation} 
then  
\begin{equation}\label{magne} 
N(\lambda ; -\Delta_{A})\; \sim \;  \frac{1}{2\pi}
\int_{\bH  } 
\mb (m) \sum_{k\in \bN } [\lambda -\frac{1}{4} -(2k+1)\mb (m)]^{0}_{+} \; dv\; . 
\end{equation} 

\end{corollary} 
The assumption (\ref{hypW}) is satisfied when 
$\Di \; \omega (\lambda )\; \sim \; \alpha \lambda ^k \ln ^j (\lambda )\ $ 
when $\lambda \to +\infty \; ,$\\
 with $\; k > 0\; ,$ or $k=0$ and $\; j > 0\; .$ 

For example this allows us to consider  magnetic fields  of the type \\
$\Di \; \mb (x,y)\; =\; \left ( \frac{x}{y} \right )^{2j} \; + \; 
g(y)\; ,$ 
with $j\; \in \; \bN^\star \; $ and $g(y)\ =\ p_1(y)\ +\ p_2(1/y)$, \\where
  $p_1(s)$ and $p_2(s)$ are,
for large $s$,   polynomial functions of order $\geq 1\ .$ 
In this case 
$\Di \; \omega (\lambda )\; \sim \; \alpha \lambda ^{\frac{1}{2j}}\ln  (\lambda )\ $
when $\lambda \to +\infty \; .$

For the proof of Theorem \ref{magn},  we will establish  some  transformations, 
  prove some technical lemmas and then use
the minimax technique on quadratic forms as in Colin de Verdi\`ere's result  to get successively a lower bound and an upper bound
for $N(\lambda ; -\Delta_{A})$.
 
\subsection{Technical  transformations}
\subsubsection{Change of variables}
  Let us consider the diffeomorphism

$\ \phi \; :\; \bR ^2\ \to \; \bH \; ,\quad (x,y)= \phi (x,t):=(x,e^t)\; $\\ 
which induces a unitary operator 
\\ 
$\widehat U\; :\; L^2(\bH ; dv)\; 
\to \; L^2(\bR ^2; dxdt)\; $
\\ 
$(\widehat U f)(x,t):=e^{-t/2}f(x, e^t)\; $ for any $f \in  L^2(\bH)$.
\\ 
$\widehat U\ $ maps $C_0^\infty(\bH) $ onto  $C_0^\infty( \bR ^2 )$ and the inverse $\widehat U^{-1}\ $
is given by
\\ 
$ (\widehat U^{-1} g) (x,y):=y^{1/2}g(x,\ln y)\; $ for each $g \in  L^2(\bR ^2).$

The quadratic form related to the operator $P(A)=-\Delta_A$ is given, for any $u \in  L^2(\bH )$, by 
$$ q(u) :=\int_{\bH ^2}\left [ | y(D_x-A_1)u|^2 + | y (D_y -A_2)u|^2
 \right ]  \frac{dxdy}{y^2}$$ 
$$=\int_{\bR ^2} \left [  | e^t(D_x-\tilde A_1)u(\phi)|^2 + 
|  e^t ( e^{-t}D_t -\tilde A_2)u(\phi)|^2 
\right ]  e^{-t}dxdt$$   
$$=\int_{\bR ^2} \left [ | e^{t/2} (D_x-\tilde A_1)u(\phi)|^2 + 
|  e^{t/2}   ( e^{-t}D_t -\tilde A_2)u(\phi)|^2 
\right ] dxdt$$ 
with $$\tilde A_i (x,t):=  A_i (x,e^t)\ \ \ ,\ i=\ 1,2\ \ .$$
After defining $w:=\widehat U u$, the preceding form becomes
$$ \widehat q(w):=\int_{\bR ^2} \left [  | e^t(D_x-\tilde A_1)w|^2 
 + |  ( e^{-t/2}D_t e^{t/2} -e^t\tilde A_2)w|^2 
 \right ] dxdt$$ 
so  $$ \widehat{P}(\tilde{A}):=\widehat U P(A) \widehat U^{-1}  =
 e^{2t}(D_x-\tilde A_1)^2+(D_t  -e^t \tilde A_2)^2 +1/4 .$$

\subsubsection{Gauge}
We  want to work with a gauge such that $A_2 =0$.
Since 
$$ \mbt \; =\; y^2\left ( \partial_xA_2 -\partial_yA_1\right )$$ 
we can  take
$$ A_1(x,y) = -\int_1^y  \frac {\mbt (x,s)}{s^2} ds\ \ \ $$
which gives
\begin{equation}\label{pot}
 \tilde A_1(x,t)~:=-\int_1^{e^t}  \frac {\mbt (x,s)}{s^2} ds
\end{equation}
$$ \MR{ and}  \ \ \widehat{P}(\tilde{A}) = e^{2t} \left[ D_x  +\int_1^{e^t}  
\frac {\mbt (x,s)}{s^2}  ds \right]^2+D_t^2 +1/4 .$$
The associated quadratic form is
 $$ \widehat{q}^{\tilde{A}}(w)=\int_{\bR ^2}\left[ | e^t(D_x-\tilde A_1) w|^2 
 + | D_t w|^2 + 1/4|w|^2\right]dx dt\ .$$

An application of the assumption (\ref{Bcontrol}) is 
  the following Lemma. 


\begin{lemma}\label{fig}
For any $a>0$ and any $\varepsilon_0>0$ small enough \\
$$(\varepsilon_0< \min\{\frac{1}{2}, \frac{1}{C_0(a+1)} \})\ ,there\ exists \ \  C_1>1\ \ such\ that \ , $$
 if $\; (x_0,y_0)\; \in \; \bH\; $ and   $ \; \mb(x_0,y_0)\; >\; 1\; ,$ 
 then 
$$\frac{1}{C_1}\ \mb (x_0,y_0)\leq\mb (x,y)\leq C_1 \ \mb (x_0,y_0)\; ;
\quad \forall \; (x,y)\in \Omega (x_0,y_0, a,\varepsilon_0)$$ 
 where $\Omega (x_0,y_0, a,\varepsilon_0) :=\{(x,y)\ /\ |x-x_0|\leq a\varepsilon_0\ y_0,\ 
|y-y_0|\leq  \varepsilon_0 y_0\}\ .$

\end{lemma} 
 The proof  comes directly from the assumption
 (\ref{Bcontrol}). 
Performing Taylor expansion ,  we get
$$|\mbt(x,y)-\mbt(x_0,y_0)|\leq (|x-x_0|+|y-y_0|)
\sup_{z\in \Omega}(|\partial_x \mbt(z)|+|\partial_y \mbt(z)|)$$
so $\displaystyle \ |\mbt(x,y)-\mbt(x_0,y_0)|\leq \varepsilon_0 C_0(a+1) y_0
\sup_{z\in \Omega}\frac{\mb(z)+1}{y}$\\ 
and the proof follows easily.

 \subsection{Technical lemmas}
  
\subsubsection{Localization in a suitable rectangle in $\bR^2$} 
 
Let $\; a_0\; >\; 1\; $ be given.

Any nonnegative constant depending only on $\; a_0\; ,$ 
will be denoted invariably $\; C\; .$

Let  $X_0 =(x_0,t_0)\; \in \; \bR^2\; $  such that
$ \mb(z_0) > 1\; ;\ (z_0=(x_0,e^{t_0})\; )\; ;\ |X_0|\; $ can be very large.  

Let us choose  $\varepsilon_0\; \in ]0, 1[\;  ,\ \varepsilon_0\; $ can be very small. 

For $\ a\in ] \frac{1}{a_0} , a_0]\; ,$ let   
\begin{equation}\label{DefK}  
K:=X_0+K_0\; ,\quad K_0\; =\; 
]-\varepsilon_0  a \frac{e^{t_0}}{2},\; \varepsilon_0 a \frac{e^{t_0}}{2} [\; 
  \times\;   
]-\frac{\varepsilon_0}{2} ,\; \frac{\varepsilon_0}{2}\; [\; .
\end{equation} 
We consider the Dirichlet operator $P_K(\tilde{A})$  on $K$ associated to the quadratic form 
 $$ \widehat{q}^{\tilde{A}}_{K}(w)=
 \int_{K}\left[ | e^t(D_x-\tilde A_1) w|^2 
 + | D_t w|^2 + 1/4|w|^2\right]dx dt\; \quad \forall \; w\; \in \; W^1_0(K)\; .$$
We are interested only by the spectrum of $P_K(\tilde{A})\; .$ 
 It is gauge invariant, 
\begin{equation}\label{gaugeInvD} 
sp(P_K(\tilde{A}))\; =\; sp(P_K(\tilde{A}+\nabla \varphi ))\; , 
\end{equation}  
so by taking $\varphi (x,t)=-\int_0^x \widetilde{A_1}(s,t_0)ds\; ,$ we can assume that 
$$
\tilde A_1(x,t)~:=-\int_{e^{t_0}}^{e^t}  \frac{\mbt (x,s)}{s^2} ds
\quad \MR{(and}\quad \tilde{A}_2=0\; )\; .$$
Let us define the magnetic potential related to a constant magnetic field 

\begin{equation}\label{defA}
A^0(x,t)=(A^0_1, 0)\; \ \MR{with}\ \quad A^0_1:=-(t-t_0)\ e^{-t_0}\ \mbt(x_0,e^{t_0})\ .
\end{equation}
We want to compare $N(\lambda ;\; P_K(\tilde{A}))$ to $N(\lambda ;\; P^0_K(A^0))\; $ 
for $\lambda >>1\; ,$ 
where $P^0_K(A^0)$ is the Dirichlet operator on $K$, 
associated to the quadratic form  
 $$ \widehat{q}^{A^0, 0}_{K}(w)=
 \int_{K}\left[ | e^{t_0}(D_x - A^0_1) w|^2 
 + | D_t w|^2 + 1/4|w|^2\right]dx dt\ \quad \forall \; w\; \in \; W^1_0(K)\; .$$
\indent 
We begin with comparing the associated magnetic potentials.

\begin{lemma}\label{fige}
Under the above assumptions there exists a constant $C\; ,$ 
 depending only on $a_0$ in (\ref{DefK}), such that for any
  $(x,t)\in  K$~: 
$$ | \tilde A_1(x,t)- A_1^0(x,t) |\; \leq \; 
 C \ \varepsilon_0^2\ e^{-t_0}\ \mb(x_0,e^{t_0})\ .$$
\end{lemma}
{\bf Proof}~: As 
\begin{equation}\label{pote}
 \tilde A_1(x,t)~:=-\int_{e^{t_0}}^{e^t}  \frac{\mbt (x,s)}{s^2} ds \; ,
\end{equation}
there exists $\tau =\tau (x) \in ]t_0,t[$ such that  
\begin{equation}\label{poe}
 \tilde A_1(x,t)=-(e^t-e^{t_0})  \frac {\mbt (x,e^{\tau})}{e^{2\tau}}
=-e^{t_0}(e^{t-t_0}-1)  \frac {\mbt (x,e^{\tau})}{e^{2\tau}}
\end{equation}
Writing $$ {\cal A}:= e^{t_0}  \frac {\mbt (x,e^{\tau})}{e^{2\tau}}-
  \frac {\mbt (x_0,e^{t_0})}{e^{t_0}} $$ we get from the definition \ref{defA} 

$$ |\tilde A_1(x,t)-  A_1^0(x,t)|\leq C |t-t_0| |{\cal A}|\ \leq  C \ \varepsilon_0\ |{\cal A}|\ .$$

But from the lemma \ref{fig},  we get the following estimate for any $(x,\tau)\in K\; :$ 
$$ |{\cal A}|=|e^{t_0}  \frac {\mbt (x,e^{\tau})}{e^{2\tau}}-
  \frac {\mbt (x_0,e^{t_0})}{e^{t_0}}|\leq 
 C \ \varepsilon_0\ e^{-t_0}\ \mb(x_0,e^{t_0})\ .$$

To see this we decompose ${\cal A} $ in 3 parts 
$${\cal A}_1= e^{t_0-2\tau} \ (\mbt (x,e^{\tau})-\mbt (x_0,e^{\tau}))$$ 
$${\cal A}_2= e^{t_0-2\tau}\  (\mbt (x_0,e^{\tau})-\mbt (x_0,e^{t_0}))$$ 
$${\cal A}_3= e^{t_0} \mbt (x_0,e^{t_0})\ (\frac{1}{e^{2\tau}}-\frac{1}{e^{2t_0}})$$ 
According to the assumption (\ref{Bcontrol}) and to the Lemma \ref{fig}  we have
 $$\ (\mbt (x,e^{\tau})-\mbt (x_0,e^{\tau}))\leq e^{t_1-t_0}\ \mb(x_0,e^{t_0})\; ,$$
 so
$$|{\cal A}_1|\leq e^{-t_0} C|x-x_0| e^{-t_0}
\ \mb(x_0,e^{t_0})\leq C a\varepsilon_0 e^{-t_0}
\ \mb(x_0,e^{t_0})\; ,$$
$$|{\cal A}_2 |
\leq e^{-t_0} C|e^{\tau}-e^{t_0}| e^{-t_0}
\ \mb(x_0,e^{t_0})\ 
 \leq C e^{-t_0} |\tau -t_0|
\ \mb(x_0,e^{t_0})\leq C\varepsilon_0 e^{-t_0}  \mb(x_0,e^{t_0}) \ ,$$ 
The third term is also bounded by the same expression
$$|{\cal A}_3| \leq  C \ \varepsilon_0\ e^{-t_0}\ \mb(x_0,e^{t_0})\ \ ,$$ 
so we finished the proof.

 \subsubsection{Quadratic forms on $ K$}

Let us define 
$$ \widehat{q}^{\tilde{A},0}_{ K}(w):=
\int_{K}  \left[ | e^{t_0}(D_x-\tilde A_1) w|^2 + 
| D_t w|^2 + 1/4|w|^2\right]dx dt\; ,
\ \forall w \in W^1_0(K)\; .$$ 
\begin{lemma}\label{fig1} There exists a constant $C$ 
depending only on $a_0$ of (\ref{DefK}),  
s.t. 
$$(1-\varepsilon_0C)\ \widehat{q}^{\tilde{A},0}_{ K}(w)\leq 
\widehat{q}^{\tilde{A}}_{ K}(w)
\leq (1+\varepsilon_0C)\widehat{q}^{\tilde{A},0}_{ K}(w)\; .$$
\end{lemma}
{\bf Proof }~: 
Write

$$ \widehat q^{\tilde{A}}_{ K}(w)=\int_{K}  \left[e^{2(t-t_0)}
| e^{t_0}(D_x-\tilde A_1)w|^2 + | D_t w|^2 dxdt| + 1/4|w|^2\right]\ dxdt$$
and use that $|t-t_0|\leq 1\;  $ in $ K\; .$

\begin{lemma}\label{fig0}
There exists a constant $C$ 
depending only on $a_0$ of (\ref{DefK}), 
  such that, for any $\tau \; \in \; ]0, 1[\; ,$ 
(with $z_0=(x_0, e^{t_0})\; )\; ,$ 
$$(1-\tau^2)\ \widehat{q}^{A^0 , 0}_{ K}(w)+(1-\frac{1}{\tau^2})\ C\ 
\varepsilon_0^4\ \mb^2(z_0) \| w\| ^2
\leq \widehat{q}^{\tilde{A},0}_{ K}(w)$$ 
$$
\leq \; (1+\tau^2) \widehat{q}^{A^0, 0}_{K}(w) +(1+\frac{1}{\tau^2})
\ C \ \varepsilon_0^4\ \mb^2(z_0)\| w\| ^2\; .$$
\end{lemma}
{\bf Proof }~: 
This is a straightforward application of lemma \ref{fige}, 
when we 
write
$$ e^{t_0}(D_x-\tilde A_1)w= e^{t_0}(D_x- A_1^0)w- 
e^{t_0}(\tilde A_1-A_1^0)w\ .$$

 \subsubsection{Spectral asymptotics for  a rectangle.} 
 
 An immediate application of Theorem \ref{rect} in the appendix is the 
 following Lemma. 
 \begin{lemma}\label{NLK} For any real $\lambda \; ,$ 
 \begin{equation}\label{NLKM} 
 N(\lambda , P^{0}_{K} (A^0) ) \; \leq \; 
 \frac{|K| b(x_0,e^{t_0})}{2\pi e^{t_0}} \sum_{k=0}^{+\infty} 
 [\lambda -\frac{1}{4} - (2k+1)b(x_0,e^{t_0}) ]_{+}^{0} \; . 
 \end{equation} 
 Moreover, there exists a constant 
 $C_0$ depending only on $a_0$ of (\ref{DefK}),  
 such that, 
 if $\; \varepsilon_{0}^{-2}/C_0 \leq b(x_0,e^{t_0}) \leq \lambda \; ,$ then 
 $\ \forall  \; \tau \; \in ]0, 1[\; ,$ 
 \begin{equation}\label{NLKm} 
 (1 - \tau )^2 
 \frac{|K| b(x_0,e^{t_0})}{2\pi e^{t_0}} \sum_{k=0}^{+\infty} 
 [\lambda -\frac{1}{4} - \frac{C_0}{(\tau \varepsilon_0)^2} - (2k+1)b(x_0,e^{t_0}) ]_{+}^{0} 
 \; \leq \; N(\lambda , P^{0}_{K} (A^0) ) \; . 
 \end{equation} 
 \end{lemma} 
 {\bf Proof.}  
Change  variables
$(x,t)\; \to \; (\xi , \theta ) =  ( e^{-t_0}(x - x_0), t-t_0)$\\ 
and apply (\ref{rectM}) to get (\ref{NLKM}), 
and (\ref{rectm}) to get (\ref{NLKm}). 

Taking into account (\ref{gaugeInvD}), Lemmas \ref{fig1} -  \ref{NLK}, 
we get the following proposition.

\begin{proposition}\label{NLKE} 
 There exists a constant $C_1\; >\; 1\; $ 
 depending only on $a_0$ of (\ref{DefK}),   
 such that for any $\; \varepsilon_0\; \in \; ]0,1/(2C_1)[\; ,$ 
 for any real $\lambda \; >\; 1 \;  $ and 
 for any $\; \eta \; \in ]0, 1/2[\; ,$  
 \begin{equation}\label{NLKEM} 
 N(\lambda , P_{K} (\tilde{A}) ) \; \leq \; 
 \frac{|K| b(x_0,e^{t_0})}{2\pi e^{t_0}} \sum_{k=0}^{+\infty} 
 [\Lambda_M(\lambda) - \frac{1}{4} - (2k+1)b(x_0,e^{t_0}) ]_{+}^{0} \; , 
 \end{equation} 
 with $\displaystyle \ \Lambda_M(\lambda )\; 
 =\; (1-\eta^2)^{-1}[\frac{\lambda}{1-\varepsilon_0 C_1} + 
 \frac{\varepsilon_0^4}{\eta^2}C_1b^2(x_0,e^{t_0}) ]\; .$

 Moreover, there exists a constant 
 $C_0$ depending only on $a_0$ of (\ref{DefK}),  
such that for any $\; \varepsilon_0\; \in \; ]0,1/(2C_1)[\; ,$ 
 for any real $\lambda \; >\; 1 \;  $ and 
 for any $\; \eta \; \in ]0, 1/2[\; ,$ 
 if $\; \varepsilon_{0}^{-2}/C_0 \leq b(x_0,e^{t_0}) \; ,$ 
 then 
 $\ \forall  \; \tau \; \in ]0, 1[\; ,$ 
 \begin{equation}\label{NLKEm} 
 (1 - \tau )^2 
 \frac{|K| b(x_0,e^{t_0})}{2\pi e^{t_0}} \sum_{k=0}^{+\infty} 
 [\Lambda_m(\lambda) - \frac{1}{4}- \frac{C_0}{(\tau \varepsilon_0)^2} - (2k+1)b(x_0,e^{t_0}) ]_{+}^{0} 
 \; \leq \; N(\lambda , P_{K} (\tilde{A}) ) \; ,  
 \end{equation} 
  with $\displaystyle \ \Lambda_m(\lambda )\; 
 =\; (1+\eta^2)^{-1}[\frac{\lambda}{1+\varepsilon_0 C_1} -  
 \frac{\varepsilon_0^4}{\eta^2}C_1b^2(x_0,e^{t_0}) ]\; .$
 \\ 
 Without the condition that 
    $\; \epsilon_{0}^{-2} /C_0 \leq  b(x_0, e^{t_0})\; ,$
  we have in the same way that 
 \begin{equation}\label{NLKEWm} 
 \frac{|K| b(x_0,e^{t_0})}{4\pi e^{t_0}}  
 [\Lambda_m(\lambda) - \frac{1}{4} - 
 C_0(\sqrt{\lambda}(1+b(x_0,e^{t_0}))^{1/2} ] 
 \; \leq \; N(\lambda , P_{K} (\tilde{A}) ) \; .  
 \end{equation}

 \end{proposition} 
 For the lower bound (\ref{NLKEWm}), use the same method as for 
 (\ref{NLKEm}), by using the lower bound in (\ref{rectW}) instead 
 of (\ref{rectm}).

 \subsection{Lower bound and upper bound for the $N(\lambda ; -\Delta_{A})$}

 \subsubsection{A partition adapted to $\mb$}\label{partL}
 
Let $\; a_0\; $ and $\; \delta_0 \; $ be given such 
\begin{equation}\label{HypAD} 
1\; < \; a_0\; \quad \MR{and} \quad \delta_0 \; 
 \in \; ]\frac{1}{3} , \frac{2}{5}[ \; . 
 \end{equation} 
 
 For any $\; \alpha \; \in \; \bZ^2 \; ,$ we denote the rectangle  
 \begin{equation}\label{DefIj} 
 K(\alpha)\; =\; ]-\frac{e^{\alpha_2}}{2} +e^{\alpha_2}\alpha_1\; , \; 
 e^{\alpha_2}\alpha_1  + \frac{e^{\alpha_2}}{2} [ \times 
 ]-\frac{1}{2} + \alpha_2\; ,\; \alpha_2 + \frac{1}{2}[\; . 
 \end{equation}   
 So $\displaystyle \; \bR^2 \; =\; \cup_{\alpha} \overline{K}(\alpha) \; $ 
  and $\; K(\alpha) \cap K(\beta) \; =\; \emptyset\; $ 
  if $\alpha \; \neq \; \beta \; .$  
 Taking into account  Lemma \ref{fig}, each $K(\alpha) $ can be 
 parted, (if necessary), into $M(\alpha )$ rectangles: 

\begin{equation}\label{partKa} 
 \overline{K}(\alpha )\; =\; \cup_{j=1}^{M(\alpha)}\overline{K_{\alpha,j}}\; , 
 \quad K_{\alpha , j}\; =\; ]-\frac{\epsilon_{\alpha ,j}e^{t_{\alpha , j}}}
{2} 
 +x_{\alpha ,j}\; ,\; x_{\alpha ,j} +  \frac{\epsilon_{\alpha ,j}
e^{t_{\alpha , j}}}{2} 
 [ \times ] -\frac{\epsilon_{\alpha ,j}}{2} + t_{\alpha , j} \; , 
 \; t_{\alpha , j} + \frac{\epsilon_{\alpha ,j}}{2} [\; , 
 \end{equation} 
 with 
 \begin{equation}\label{partKaE} 
 \frac{1}{a_0(1+\mb ^{\delta_0}(x_{\alpha ,j}, e^{t_{\alpha ,j}})\; )} 
 \; \leq \; \epsilon_{\alpha , j}\; \leq \; 
 \frac{a_0}{(1+\mb ^{\delta_0}(x_{\alpha ,j}, e^{t_{\alpha ,j}})\; )} 
\; ,
\end{equation} 
and such that $\displaystyle \;  K_{\alpha , k} \cap K_{\alpha , j}\; 
=\; \emptyset \; $ if $\; k\neq j\; .$ \\ 

We will denote $\displaystyle \ \Gamma \; =\; \{ (\alpha , j)\; ;\ \alpha \in 
\bZ^2\; ,\ j\in \{ 1, \ldots , M(\alpha)\} \} \; ,$\\ 
$\displaystyle \ X_\gamma \; =\; (x_\gamma , t_\gamma )\; $ the center 
of $\; K_\gamma \; ,\ ( \gamma \; \in \; \Gamma )\; ,$ 
and $\; z_\gamma \; =\; (x_\gamma , e^{t_\gamma } )\; .$

 \subsubsection{The lower bound estimate}
 
 \begin{proposition}\label{minNL} Under the assumptions 
 of Theorem \ref{compactR} and on $\; \delta_0\; $ in (\ref{HypAD}), 
 there exists a constant $\; C_0\; > \; 0\; $ such that 
 \begin{equation}\label{minNLE} 
 \frac{1}{2\pi} \int_{\bH} (1-\frac{C_0}{(\mb (m)+1)^{(2-5\delta_0)/2}}) 
 \mb (m) \sum_{k=0}^{+\infty} [\lambda (1-C_0\lambda^{-3\delta_0+1})-\frac{1}{4}  
 - (2k+1) \mb (m)]_{+}^{0} \; dv \; 
 \leq \; N(\lambda , -\Delta_A)\; . 
 \end{equation} 
 
\end{proposition} 
{\bf Proof.} Any constant  depending only on the assumptions will 
be denoted invariably by $C\; .$ 

 As $\displaystyle \; \bR^2\; =\; \bigcup_{\gamma \in \Gamma } 
\overline{K}_\gamma \; ,$ and $ K_\gamma \cap K_\rho = \emptyset \; $ 
if $\gamma \neq \rho \; ,$\\ 
we get that $\displaystyle \; 
\sum_{\gamma \in \Gamma} N(\lambda , P_{K_\gamma}(\tilde{A}) ) \; 
\leq \; N(\lambda , -\Delta_A ) \; .$\\ 
For large $|\gamma | \; ,$ we use the lower bound estimate 
(\ref{NLKEm}) ,  
with 

$$ \eta ^2= \epsilon_\gamma = b^{-\delta_0}(z_\gamma)\; \  
\MR{and}\  \; \tau =b^{-(5\delta_0 -2)/2}(z_0)\; ,
 $$ 
and also the fact that on $\displaystyle   \; K_\gamma \; ,\quad 
|e^{-t}\mb (x,e^t)  -e^{-t_\gamma} \mb (z_\gamma ) | \leq \epsilon_\gamma C\; .$\\ 
For small $|\gamma |\; ,$ (even the $\gamma $ such that 
$b(z_\gamma ) \leq \lambda ^{1 -2(3\delta_0 -1)}\; )\; ,$ 
we use (\ref{NLKEWm}) instead of (\ref{NLKEm}), taking into account 
the Remark \ref{compar}.   
The lower bound (\ref{minNLE})  comes easily.

\subsubsection{The upper bound estimate}
 
 \begin{proposition}\label{majNL} Under the assumptions 
 of Theorem \ref{compactR} and on $\; \delta_0\; $ in (\ref{HypAD}), 
 there exists a constant $\; C_0\; > \; 0\; $ such that 
 \begin{equation}\label{majNLE} 
 N(\lambda , -\Delta_A) \; \leq \;  
 \end{equation} 
 $$ \frac{1}{2\pi} \int_{\bH} (1+\frac{C_0}{(\mb (m)+1)^{(2-5\delta_0)/2}}) 
 \mb (m) \sum_{k=0}^{+\infty} [\lambda (1+C_0\lambda^{-3\delta_0+1})-\frac{1}{4} 
 - (2k+1) \mb (m)]_{+}^{0} \; dv \; 
 \; . $$
 
\end{proposition} 
{\bf Proof.} 
 Any constant  depending only on the assumptions will 
be denoted invariably by $C\; .$ 

 We keep the partition and the notation used 
in the lower bound. 

We consider the covering of $\bR^2$ by open rectangles: 
$$ \bR^2\; =\; \bigcup_{\gamma \in \Gamma } 
\cK_\gamma \; ,\quad \cK_\gamma \; =\; X_\gamma +(1/\tau_\gamma ) 
(K_\gamma -X_\gamma)\; ,$$ 

with $\; \tau_\gamma \; \in \; b^{-(5\delta_0-2)/2}(z_\gamma)[a_{0}^{-1} , a_0]\; .$ 
Then there exists a partition of unity 
$(\chi_\gamma (x,t))\; $ satisfying 
\begin{equation}\label{PropPart} 
\left \{  \begin{array}{c} \sum_\gamma \chi^2_\gamma =1 \\ 
\MR{support}(\chi_\gamma) \subset \cK_\gamma \\ 
|D_x\chi_\gamma |\leq C/(e^{t_\gamma} \epsilon_\gamma \tau_\gamma )\\ 
|D_t\chi_\gamma | \leq C /( \epsilon_\gamma \tau_\gamma ) 
\end{array} 
\right \} 
\end{equation}

We write $\displaystyle \; \hat{q}^{\tilde{A}}(w)\; =\; \sum_{\gamma } 
\left [ \; \hat{q}_{\cK_\gamma}^{\tilde{A}} (\chi_\gamma w) - 
\int_{\cK_\gamma } V |\chi_\gamma w|^2 dxdt\; \right ]\; ,$\\ 
with $\displaystyle \; V(x,t)=\sum_{\gamma} [|D_x\chi_\gamma (x,t)|^2 
+|D_t\chi_\gamma (x,t)|^2]\; .$ \\ 
Thus on 
$\cK_\gamma \; ,\ V(x,t) \leq C/(\epsilon_\gamma \tau_\gamma )^2\; ,$ 
and it follows easily from the min-max principle that 
\begin{equation}\label{SomMaj} 
N(\lambda , -\Delta_A)\; \leq \; 
\sum_\gamma N(\lambda +\frac{C}{(\epsilon_\gamma \tau_\gamma )^2} ,\; 
P_{\cK_\gamma } (\tilde{A}) )\; .
\end{equation} 
Then we get  (\ref{majNLE}) from (\ref{SomMaj}), as for (\ref{minNLE}), 
but using only (\ref{NLKEM}), (instead of (\ref{NLKEm}) and (\ref{NLKEWm}) ).

\begin{center} 
{\bf APPENDIX} 
\end{center}

\appendix
 
\section{Constant magnetic laplacian on the flat plane}

\subsection{The density of states for the euclidian constant magnetic field}

Let us  consider on $L^2(\bR^2)$ the Schr\"odinger 
operator with a constant magnetic field  
$\Di \; H_0=(D_x - b \frac{y}{2})^2 + (D_y +b \frac{x}{2})^2\; ;
\quad  ( b\; >\; 0\;$ is a constant). \\ 
The density of states of $H_0,\ \cN (\lambda , H_0 )\; ,$ is defined, 
(see \cite{D-I-M}), by 
\begin{equation}\label{DDst} 
\cN ( \lambda , H_0 )\; =\; 
\lim_{R\to \infty} \frac{N(\lambda ,H_{0}^{\Omega_R})}{|\Omega_R |}\; ; 
\end{equation} 
$\Omega_R$ is any bounded open domain of $\bR^2\; ,$ with Lipschitz 
boundary, containing  $(]-\frac{R}{2},\; \frac{R}{2}[)^2\; ,$  
 
and $H_{0}^{\Omega_R}$ is any self-adjoint operator on 
$L^2(\Omega_R)$ associated to the quadratic form of 
$H_0\; ,$ with domain included in 
the Sobolev space $W^1(\Omega_R)\; .$

\begin{theorem}\label{Col2D} The Colin de Verdi\`ere formula holds 
for any $\lambda > 0\; :$ 
\begin{equation}\label{Col2DF} 
\cN (\lambda , H_0)\; =\; \frac{b}{2\pi } \sharp \{ n\in \bN \; ;\ (2n+1)b < \lambda \} \; . 
\end{equation} 
\end{theorem} 
{\bf Proof:} Its comes easily from THEOREME 1.6 of \cite{Dem}. 

Let us  sketch a proof.

By scaling and dividing $\lambda $ by $b$, we need only to establish 
the formula when $\; b=1\; .$

We take $\Omega_R=(]-\frac{R}{2}, \frac{R}{2}[)^2\; ,$  the Dirichlet 
boundary conditions on $x=\pm \frac{R}{2} $ and 
the Floquet conditions: $e^{ixy/2}u(x,y)$ is $R$-periodic in $y\; .$ \\ 
As $u\to ue^{ixy/2}$ is a unitary operator, by performing this 
 gauge transform, $H_{0}^{\Omega_R}$ becomes 
$H_0=D_x^2 + (D_y +x)^2\; ,$ for the Dirichlet boundary 
conditions on $x=\pm \frac{R}{2}$ and the periodic ones on 
$y=\pm \frac{R}{2}\; .$ 

$H_{0}^{\Omega_R}$ and  $H_0$ have the same spectrum.

Using discret Fourier expansion, we get that 
$\Di \; N(\lambda , H_{0}^{\Omega_R})=\sum_{k\in \bZ} 
N(\lambda , H_{k,R})\; ,$\\ 
where $H_{k,R}$ is the Dirichlet operator on 
$I_R=]-\frac{R}{2} , \frac{R}{2}[\; ,$ 
associated to the harmonic oscillator 
$\; D_x^2 + (\frac{2k\pi}{R}+x)^2\; .$
 
As $N(\lambda , H_{k,R})=0$ when $|k|>\frac{1}{2\pi}(R\sqrt{\lambda} 
+\frac{R^2}{2})\; ,$
we get
$$\Di \cN (\lambda , H_0) \leq  
\sharp \{ n\in \bN ;\ 2n+1 < \lambda \} \times 
\lim_{R\to \infty} 
\frac{1}{\pi R^2}(R\sqrt{\lambda} +\frac{R^2}{2})\; ,$$
\\ 
or equivalently $\ \Di \cN (\lambda , H_0) \; \leq 
  \; \frac{1}{2\pi } \sharp \{ n\in \bN ;\ 2n+1 < \lambda \} \; .$

Now, for any fixed $\epsilon \in ]0,1[\; ,$ ( for example 
$\epsilon =1/\sqrt{R}\;) ,$  
and for any $k$ such that $|k|\leq (1-\epsilon )\frac{R^2}{4\pi}\; ,$  the exponential decreasing of the eigenfunctions of 
the harmonic oscillator on $\bR^2\; $ leads to 

$\Di N(\lambda , H_{k,R})\; \geq \sharp 
\{ n\in \bN ;\ 2n+1 < \lambda -\frac{C_\lambda}{\epsilon^2R^2} \} \; ,$\\ 
where $C_\lambda $ depends only on $\lambda \; .$ 

To see this, with  $\epsilon $ chosen as previously   and
 $R > > \lambda +1,$ 

 just use the fact that, for any 
 $\;  u\; \in \; \chi (4x/\sqrt{R})E_{\lambda}(H_{k,\infty})[L^2(\bR)]
 \; , $ 
 $$ \int_{-R/2}^{R/2}H_{k,R}\ u\overline{u}\ dx \; 
 \leq \; (\lambda +C/R)  \int_{-R/2}^{R/2} |u|^2dx\; .$$ 
 $H_{k,\infty}$  denotes the harmonic oscillator 
 $\; D_x^2 + (\frac{2k\pi}{R}+x)^2\; $ on $L^2(\bR )\; ,$  
 $\chi $ is a cut-off function, supported in $[-1,1]$ and equal to $1$ in 
 $[-1/2,1/2]\; $ ,\\ 
 and $\ \ E_\lambda (H_{k,\infty})\; $ denotes the spectral 
 projection on $]-\infty ,\lambda [\;$ of the self-adjoint operator $H_{k,\infty}$.

 Then, with the same $\epsilon $   
and using the left-hand side continuity of the function 
$\lambda \to \sharp \{ n\in \bN ;\ 2n+1 < \lambda \} \; ,$   
 we get also that 
 $$ \cN (\lambda , H_0)\; \geq \; 
 \frac{1}{2\pi } \sharp \{ n\in \bN ;\ 2n+1 < \lambda \} \; .$$

\subsection{Eigenvalues estimate in the euclidian rectangle for a constant 
magnetic field}

Let us  consider the Dirichlet problem 
$H_{D,b}^{\Omega_R}$ associated to the Schr\"odinger 
operator with a constant magnetic field 
$\Di \; H_0=(D_x - b \frac{y}{2})^2 + (D_y +b \frac{x}{2})^2\; ;$\\ 
$  ( b\; >\; 0\;$ is a constant),  
in a rectangle $\Omega_R=]-\frac{R_1}{2} ,\; \frac{R_1}{2}[\times 
]-\frac{R_2}{2} ,\; \frac{R_2}{2}[\; ;\ R=(R_1,R_2)\; \in \; (\bR_{+}^{\star})^2\; .$

\begin{theorem}\label{rect} The Colin de Verdi\`ere upper bound holds 
for any $\lambda > 1\; :$ 
\begin{equation}\label{rectM} 
N (\lambda , H_{D,b}^{\Omega_R})\; \leq\;  \frac{b | \Omega_R |}{2\pi } 
\sharp \{ n\in \bN \; ;\ (2n+1)b < \lambda \} \; . 
\end{equation}

For the lower bound,  we will need to precise  Colin de Verdi\`ere's one as follows. \\ 
There exists a constant $C_0>0$ s.t., if $0<b<\lambda $ and 
$1\leq \sqrt{b}\; \min R_j\; ,$ then $\forall \; \epsilon \; \in ]0,1]\; ,$  
\begin{equation}\label{rectm}
 (1-\epsilon )^2\frac{b | \Omega_R |}{2\pi } 
\sharp \{ n\in \bN \; ;\ (2n+1)b < \lambda -\frac{C_0}{(\epsilon \min R_i )^2} \} \;  
\leq \; 
N (\lambda , H_{D,b}^{\Omega_R})\; . 
\end{equation}

The classical Weyl estimate is the following. 
There exists a constant $C_0>0$ such that, if $0\leq b \leq \lambda $ 
and $(C_0\sqrt{b})^{-1}\leq R_j\; $ for $j=1,\ 2\; ,$ then 
\begin{equation}\label{rectW} 
\frac{|\Omega_R |}{4\pi}\lambda [1 - C_0\frac{\sqrt{b}}{\sqrt{\lambda}} ]
\; \leq \; N(\lambda , H_{D,b}^{\Omega_R} )
\; \leq \frac{|\Omega_R|}{4\pi}\lambda [1 + C_0\frac{\sqrt{b}}{\sqrt{\lambda}} ]
\; . 
\end{equation}

\end{theorem}    
{\bf Proof:} The upper bound  (\ref{rectM}) and the lower bound 
(\ref{rectm}) come  
from the density of state using the same proof as in 
 Colin de Verdi\`ere paper \cite{Col}.

We sketch the proof of the lower bound (\ref{rectm}). 

 We set :~$ R(b)=R\sqrt{b}\; .$ By scaling, we change 
 
$\Omega_R\; ,\ b$ and $\lambda$ into $\Omega_{R(b)}\; ,\ 1$ 
and $\lambda/b\;   .$  

Then we take a large rectangle 
$\Di \Omega_{R(b),M,\epsilon }\; 
=\; \bigcup_{j=1}^{M}
\overline{\Omega (j,\epsilon )}\; $ 
where 

$\Omega (j,\epsilon )=z_j+(1-\epsilon)\Omega_{R(b)}$ 
are open rectangles with center $z_j$  such that 

$\Di \Omega (j,\epsilon ) \bigcap \Omega (k,\epsilon ) 
= \emptyset $ if $j\neq k\; .$ \\ 
We consider the large rectangle $\Di \Omega_{R(b),M }\; 
=\; \bigcup_{j=1}^{M}
 \Omega (j)\; $ 
where $\Omega (j)=z_j+\Omega_{R(b)}.$ 
 
So there exists a constant $C_0$ and a
partition of unity $(\chi_j)$ s.t. 
$$\MR{support}(\chi_j)\; \subset \; \overline{\Omega (j)}\; ,\quad 
\sum_{j=1}^{M}\chi_j^2(z)=1\ \MR{on}\ 
\Omega_{R(b),M,\epsilon }\; \quad \MR{and}\quad 
|\nabla \chi_j|\leq \frac{C_0}{\epsilon \min R_k(b)}\; .$$ 
We can write, for any $\Di \ \; u\; \in \; W_0^1(\Omega_{R(b),M,\epsilon })\; ,$ 
$$  
\int_{\Omega_{R(b),M,\epsilon }}   |(D-A_0)u|^2 dxdy
\; =\; \sum_{j=1}^{M}\int_{\Omega (j)} 
[ |(A-A_0)\chi_ju|^2 -V|\chi_j u|^2]\; dxdy\; ,$$
where $D-A_0=(D_x-\frac{y}{2}, D_y+\frac{x}{2})\; $ 
and $\Di V(z)=\sum_{\ell =1}^{M} |\nabla \chi_\ell (z)|^2\; ;$  \\ 
so  we get 
that for any real $\mu \; ,$ 
$$N(\mu , H_{D,1}^{\Omega_{R(b),M,\epsilon }} )\; 
\leq \; M\times N(\mu + (\frac{C_0}{\epsilon \min R_k(b)})^2 ,\; 
H_{D,1}^{\Omega_{R(b)}} )\; .$$
By the density of states formula, we have 
$$ \lim_{M\to \infty} \frac{N(\mu , H_{D,1}^{\Omega_{R(b),M,\epsilon }} )}{M}
\; =\; (1-\epsilon)^2\frac{bR_1R_2}{2\pi} 
\sharp \{ n\in \bN \; ;\ 2n+1 < \mu \} \; ,$$ and
we get the lower bound (\ref{rectm}) by taking 
$\mu =\frac{\lambda}{b} -  (\frac{C_0}{\epsilon \min R_k(b)})^2\; .$

For the proof of the classical Weyl estimates (\ref{rectW}), by scaling, we change 
$\Omega_R\; ,\ b$ and $\lambda$ into $\Omega_{\sqrt{b}R}\; ,\ 1$ 
and $\lambda/b\; .$ \\ 
Then we take a partition $\Di \Omega_{\sqrt{b}R}\; =\; \bigcup_{j=1}^{M}
\overline{\Omega (j)}\; $ 
where $\Omega (j)$ are open rectangles with sides in $[1/2,1]$ such that 
$\Di \Omega (j) \bigcap \Omega (k) = \emptyset $ if $j\neq k\; .$ 

So $\Di \sum_{j=1}^{M} N(\frac{\lambda}{b} , H_{D,1}^{\Omega (j)}) 
\; \leq \; N(\lambda , H_{D,b}^{\Omega_R})\; .$   

We change gauge in each $\Omega (j)$ in order to consider 
$H_{D,1}^{\Omega (j)}$ as the operator 

$(D-A(j))^2=(D_x -\frac{y-y_j}{2})^2+(D_y + \frac{x-x_j}{2})^2\; ,$ 
where $(x_j,y_j)$ is the center of $\Omega (j)\; .$ 

Now, it is easy to get the uniform Weyl formula:

 $\ \exists C_0\; >\; 0\quad \MR{s.t.} \quad \forall \; j+1,\ldots ,M\; ,$  
$$ \frac{|\Omega (j) |}{4\pi} \frac{\lambda}{b}
[1-C_0\frac{\sqrt{b}}{\sqrt{\lambda}}] 
\leq N(\frac{\lambda}{b} , H_{D,1}^{\Omega (j)}) \leq 
\frac{|\Omega (j) |}{4\pi} \frac{\lambda}{b}
[1+C_0\frac{\sqrt{b}}{\sqrt{\lambda}}]  \; .$$
To be convinced, see that $\exists \; C_0 >0$ s.t. $\forall 
\; \tau \; \in ]0,1]\; ,\ \forall \; u\; \in 
W_0^1(\Omega (j)\;) ,$  
$$(1-\tau^2)\| \nabla u \|^2 
- \frac{C_0}{\tau^2}\| u\|^2 
\leq \| (D-A(j))u\|^2  
\leq (1+\tau^2)\| \nabla u \|^2 
+\frac{C_0}{\tau^2}\| u\|^2\; ,$$ 
and take $\tau = \frac{\sqrt{b}}{\sqrt{\lambda}}\; .$ 

So we get the lower bound of (\ref{rectW}). 
 We get in the same way the upper bound by considering the Neumann 
 operators $H_{N,1}^{\Omega (j)}$ instead of the Dirichlet ones 
 $H_{D,1}^{\Omega (j)}\; .$

\begin{remark}\label{compar} As in Theorem \ref{rect},  
$$\frac{\lambda -b}{2} \; \leq \; \sharp \;  \{ n\in \bN \; ; \ 
(2n+1)b < \lambda \} \; \leq \; \frac{\lambda +b}{2}\; ,$$
so the upper bound (\ref{rectM}) is sharp compared to the one in 
(\ref{rectW}). The lower bound (\ref{rectm}) is sharp, compared to the 
one in (\ref{rectW}), when $\; \epsilon \; <\; \sqrt{(b/\lambda)}\; .$ 
\end{remark}

\end{document}